\documentstyle[graphicx, 12pt]{article}
\begin{document}

\small
\hoffset=-1truecm
\voffset=-2truecm
\title{\bf Description of phase transition in a black hole
with conformal anomaly in the Ehrenfest's scheme}
\author{Jingyun Man \hspace {1cm} Hongbo Cheng\footnote {E-mail address: hbcheng@ecust.edu.cn}\\
Department of Physics, East China University of Science and
Technology,\\ Shanghai 200237, China\\
The Shanghai Key Laboratory of Astrophysics, Shanghai 200234,
China}

\date{}
\maketitle

\begin{abstract}
We make use of the Ehrenfest's equations to explore the phase
transition of a black hole with conformal anomaly. The first-order
phase transition is ruled out because no discontinuity appears in
entropy of the black holes. We find that the phase transition of such
black holes belongs to the second order subject to the Ehrenfest's
equations. Further we also show that the second-order phase
transition will not happen for the black holes without conformal
anomaly.
\end{abstract}
\vspace{7cm} \hspace{1cm} PACS number(s): 04.70.Bw, 14.80.Hv\\
Keywords: Black hole, Phase transition, Ehrenfest's scheme

\newpage

\noindent \textbf{I.\hspace{0.4cm}Introduction}

The Einstein-Gauss-Bonnet gravity including various spacetimes is
of considerable interest motivated by developments in the string
theory. The theory is a special case of Lovelock's theory of
gravitation [1]. The Gauss-Bonnet term arising naturally in the
low-energy limit of heterotic superstring theory is the first and
dominating quantum correction to classical general relativity.
This term appears as quadratic in the curvature of the spacetime
in the Lagrangian. The Gauss-Bonnet term regularizes the metric
and modifies the Friedmann equation. The estimation on the
Gauss-Bonnet effect has been performed [2-8]. In four dimensions,
Gauss-Bonnet term reduces to a topological invariant unless the
surface term is considered. In higher dimensions, it is a strong
correction which encodes the property of field equation that only
contains the second derivative of metric. Many efforts have been
made to study the thermodynamics of black holes in Gauss-Bonnet
gravity [4,23-28]. According to the first law of thermodynamics,
they find the entropy of Gauss-Bonnet black holes in high
dimensions is not directly proportional to the area of event
horizon like Bekenstein-Hawking entropy, but also carry a
Gauss-Bonnet term with coupling parameter. As an important concept
in quantum field theory in curved spacetimes, the conformal
anomaly is inserted into the energy-momentum tensor within the
gravity and cosmology. The conformal anomaly hires the
Gauss-Bonnet term. We adopt the four-dimension entropy with a
logarithmic term proposed by Cai etc. [7], and further discuss the
character of phase transition of the black holes with conformal
anomaly. It is significant to probe its influence in different
directions.

Recently a lot of attentions were paid to the thermodynamic
quantities and phase transitions on various black holes due to the
area law of the black hole entropy provoked by Bekenstein and
Hawking [9-11]. The black holes can be thought as thermodynamic
system. The works on phase transitions of black holes in the frame
of semiclassical gravity were listed [12]. The thermodynamic
characteristics of modified Schwarzschild black hole have been
researched in the Refs. [7, 13]. The thermodynamic phase transition
in Born-Infeld-anti de Sitter black holes were discussed in virtue
of various ways [14, 15]. The phase transition of the
quantum-corrected Schwarzschild black hole was investigated, which
fosters the research on the quantum-mechanical aspects of
thermodynamic behaviours [16]. The thermodynamic quantities of a
black hole involving an $f(R)$ global monopole were also evaluated
[17]. In the process of research on the phase transition of black
holes, a new idea based on the Clausius-Clapeyron scheme or the
Ehrenfest's scheme was put forward by Banerjee et. al. [18, 19].
Further the phase transition in Reissner-Nordstrom-AdS black holes
and Kerr-AdS black holes were discussed with the help of
Ehrenfest's equations [19-21].

In this paper we plan to study the phase transition of static and
spherically symmetric black holes with conformal anomaly in the
Ehrenfest's scheme classifying the phase transitions as the first
order or higher order transition. The thermodynamic quantities and
phase transition of black holes with Gauss-Bonnet corrections were
discussed [7, 22]. It is significant to elaborate how the
Gauss-Bonnet effect control the black holes phase transition. We
should understand which kind of phase transition happens for the
black hole while how their thermodynamic quantities behave. To our
knowledge, little contribution is made to estimate the influence
from Gauss-Bonnet term on the transition. We wish to make use of
Ehrenfest's equations to describe the phase transition of the
spherically symmetric black holes whose effective energy-momentum
tensor contains the Gauss-Bonnet term. We follow the procedure of
Refs. [18, 19] to study the evolution of this kind of black holes.
First of all the exact soluble metric with the quantum black
reaction will be introduced. We analyze the nature of the phase
transition of the Gauss-Bonnet corrected black hole by means of
Ehrenfest's scheme to show the relation between the phase
transition and the Gauss-Bonnet term. We derive the thermodynamic
variables such as the heat capacity etc. and discuss their
singularities at the critical point to exhibit which order the
phase transition of the black hole is. Finally the arguments will
be listed.

\vspace{0.8cm} \noindent \textbf{II.\hspace{0.4cm}Thermodynamics of black hole with conformal anomaly}

When the Gauss-Bonnet term is inserted into the effective
energy-momentum tensor, the static and spherically symmetric
metrics as solutions to the Einstein equation are found as [1],

\begin{equation}
ds^{2}=f(r)dt^{2}-\frac{dr^{2}}{f(r)}-r^{2}(d\theta^{2}+
\sin^{2}\theta d\varphi^{2}),
\end{equation}

\noindent where

\begin{equation}
f(r)=1-\frac{r^{2}}{4\alpha}(1-\sqrt{1-\frac{16\alpha M}{r^{3}}
+\frac{8\alpha Q^{2}}{r^{4}}}).
\end{equation}

\noindent Here $M$ and $Q$ are integration constants, and the
coefficient $\alpha$ is positive. We set the Newton constant
$G_{N}=1$. The asymptotic behaviour of this line element is given
by,

\begin{equation}
f(r)=1-\frac{2M}{r}+\frac{Q^{2}}{r^{2}}+O(\alpha, r^{-2}),
\end{equation}

\noindent which is the form of Reissner-Nordstrom black holes with
the mass $M$ and the electric charge $Q$ asymptotically. The outer
horizon is a root of metric (2) like $f(r_{+})=0$, then the mass
of the black hole can be written as,

\begin{equation}
M=\frac{r_{+}}{2}+\frac{Q^{2}-2\alpha}{2r_{+}}.
\end{equation}

\noindent The correlation between mass $M$ of a black hole with
conformal anomaly and the outer horizon $r_{+}$ is plotted in
Figure 1. When $r_{+}=\sqrt{Q^{2}-2\alpha}$, the mass of black
hole achieves the minimum. The mass function decreases while
$0<r_{+}<\sqrt{Q^{2}-2\alpha}$. According to Ref. [7], the
modified entropy is,

\begin{equation}
S=\pi r_{+}^{2}-4\pi\alpha\ln r_{+}^{2}.
\end{equation}

\noindent The first law of black hole thermodynamics is [7],

\begin{equation}
dM=T_{H}dS+\Phi dQ,
\end{equation}

\noindent where $T_{H}$ is the Hawking temperature denoted as
[11],

\begin{eqnarray}
T_{H}=\frac{f'(r_{+})}{4\pi}\hspace{1cm}\nonumber\\
=\frac{1}{4\pi}\frac{r_{+}^{2}+2\alpha-Q^{2}}
{r_{+}(r_{+}^{2}-4\alpha)},
\end{eqnarray}

\noindent and $\Phi$ is the potential difference between the
horizon and the infinity like [18-21],

\begin{equation}
\phi=\frac{Q}{r_{+}},
\end{equation}

\noindent and the heat capacity at constant potential of the black
hole is defined as [18-21],

\begin{eqnarray}
C_{\phi}=T_{H}(\frac{\partial S}{\partial T_{H}})_{\phi}\hspace{4cm}\nonumber\\
=-\frac{2\pi(r_{+}^{2}-4\alpha)^{2}(r_{+}^{2}-Q^{2}+2\alpha)}
{r_{+}^{4}-(Q^{2}-10\alpha)r_{+}^{2}-(4\alpha Q^{2}+8\alpha^{2})}.
\end{eqnarray}

\noindent Having found the roots of the denominator of the heat
capacity, we obtain the points of phase transition as follow,

\begin{equation}
r_{c}=\frac{\sqrt{2}}{2}\sqrt{Q^{2}-10\alpha +\sqrt{Q^{4}-4\alpha
Q^{2}+132\alpha^{2}}}.
\end{equation}

\noindent If the horizon of black hole is equal to $r_{c}$, the
phase transition will emerge. For a stable black hole with a
positive heat capacity, the event horizon radius is restricted as
$r_{c}<r_{+}<\sqrt{Q^{2}-2\alpha}$. According to expression (9),
the dependence of heat capacity on the horizon is shown
graphically in Figure 2. At points of phase transition the heat
capacity is not continuous while the divergence generates. It is
clear that the heat capacity is discontinuous at $r_{+}=r_{c}$
which is the phase transition point. At this \emph{position},
$C_{\phi}$ flips from negative infinity to positive infinity. It
indicates that black hole transform from unstable phase 1 where
the heat capacity is negative into stable phase 2 where $C_{\phi}$
is positive at the critical point. In addition to Figure 1, the
slope of $M(r_{+})$ is negative for $r_{+}<\sqrt{Q^{2}-2\alpha}$
where the transition point is included. Therefore, the phase
transition of a black hole with conformal anomaly can be described
as an unstable black hole with larger mass \emph{and smaller
radius $0<r_{+}<r_{c}$} in phase 1 turns to a stable black hole
with smaller mass and larger radius
$r_{c}<r_{+}<\sqrt{Q^2}-2\alpha$ in phase 2.

According to Ref. [18-21] we also plot the case for
Reissner-Nordstrom black hole in Figure 2. The correlation between
heat capacity and entropy for Reissner-Nordstrom black hole can be
easily expressed as $C_{\phi}=-2\pi r_{+}^{2}$, where $r_{+}$ is
the Reissner-Nordstrom black hole radius satisfied with the
equation $r_{+}^{2}-2Mr_{+}+Q^{2}=0$. The expression of heat
capacity is completely different from the case for black hole with
conformal anomaly, and it can be obtained from (9) by taking
$\alpha =0$. It can be checked that higher order derivation of
$C_{\phi}$ for Reissner-Nordstrom black hole is constant, which
means that this kind of black hole will not perform phase change
in any order.

\vspace{0.8cm} \noindent \textbf{III.\hspace{0.4cm}Description
of a black hole with conformal anomaly in the Ehrenfest's scheme}

In the standard thermodynamics, the phase transition is
fundamental phenomena for a  thermodynamical system. A phase
transition means that a discontinuity of a state space variable
occurs. That which kind of state variables have a discontinuity at
the critical points determines which kind of phase transitions.
Black hole thermodynamics is based on Hawking radiation and we can
treat a black hole as a thermodynamic system. Now we use the
classification of phase transitions proposed by Ehrenfest [18-21]
to discriminate which order of phase transition does a black hole
with conformal anomaly occur. Specifically first-order phase
transitions involve a discontinuous entropy which is a first
differential of Gibbs free energy. While in second-order phase
transitions the entropy of black holes does not show a
discontinuity, but the heat capacity and the other second
differentials of G as the analogy of isothermal compressibility
etc. do. Moreover, if the conformal anomaly leads a second-order
phase transition as we demonstrate later, the Ehrenfest relations
should be confirmed at phase transition points.

First of all we focus on the Hawking temperature (7) and the
relationship between the temperature and mass of black hole is
plotted in Figure 3. We can see that the shape of these curves for
black hole with conformal anomaly are different from
Reissner-Nordstrom black hole. The shapes of curves for black hole
involving conformal anomaly are similar. It is evident that the
curves are continuous, which means that the evolution of the black
holes with conformal anomaly does not belong to the first order
phase transition.

Next we are going to check the existence of the second order phase
transition describing how the black holes with conformal anomaly
undergo. The first and second Ehrenfest's equations for black
holes are [18-21],

\begin{equation}
-\frac{d\phi}{dT_{H}}=\frac{1}{T_{H}Q}\frac{C_{\phi 2}-C_{\phi
1}}{\beta_{2}-\beta_{1}},
\end{equation}

\begin{equation}
-\frac{d\phi}{dT_{H}}=\frac{\beta_{2}-\beta_{1}}{\kappa_{2}-\kappa_{1}},
\end{equation}

\noindent where the subscripts 1 and 2 stand for phase 1 and 2
respectively.

We discuss the analog of volume expression  coefficient $\beta$
and the analog of isothermal compressibility $\kappa$ by means of
Eq. (7) and (8),

\begin{equation}
\beta=-\frac{4\pi r_{+}(r_{+}^{2}-4\alpha)^{2}}
{r_{+}^{4}-(Q^{2}-10\alpha)r_{+}^{2}-4\alpha(Q^{2}+2\alpha)},
\end{equation}

\begin{equation}
\kappa=\frac{r_{+}}{Q}\frac{r_{+}^{4}-(3Q^{2}-10\alpha)r_{+}^{2}
+4\alpha(Q^{2}-2\alpha)}{r_{+}^{4}-(Q^{2}-10\alpha)r_{+}^{2}
-4\alpha(Q^{2}+2\alpha)}.
\end{equation}

\noindent The new variables $\beta$ correspond volume expansivity
and $\kappa$ correspond isothermal compressibility can be seen in
Figure 4 and Figure 5 respectively. It is interesting that both
the two thermodynamic quantities are discontinuous at the points
of phase transition. The natures of variables $C_{\Phi}$, $\beta$
and $\kappa$ predict that higher order phase transition in the
black holes will emerge. For Reissner-Nordstrom black hole, volume
expansivity $\beta$ and isothermal compressibility $\kappa$ are
$\beta =-\frac{4\pi r_{+}^{3}}{r_{+}^{2}-Q^{2}}$, $\kappa
=\frac{r_{+}}{Q}\frac{r_{+}^{2}-3Q^{2}}{r_{+}^{2}-Q^{2}}$, which
are also plotted in Figure 4 and 5. They are divergent at
$r_{+}=Q$. These two quantities diverse contrast to the behaviour
of $C_{\phi}$ for Reissner-Nordstrom black hole.

Now we combine the
Ehrenfest's equation (11), heat capacity (9) and the analog of
volume expansion coefficient (13) at phase transition points (10)
to obtain,

\begin{equation}
\frac{1}{TQ}\frac{C_{\phi 2}-C_{\phi 1}}{\beta_{2}-\beta_{1}}
|_{r_{+}=r_{c}}=\frac{2\pi(r_{c}^{2}-4\alpha)}{Q}.
\end{equation}

\noindent Having taken into account the Ehrenfest's equation (12),
the analog of volume expansion coefficient (13) and the analog of
isothermal compressibility (14) at the critical points (10), we
find

\begin{equation}
\frac{\beta_{2}-\beta_{1}}{\kappa_{2}-\kappa_{1}}|_{r_{+}=r_{c}}
=\frac{2\pi(r_{c}^{2}-4\alpha)}{Q}.
\end{equation}

\noindent Although the variables $C_{\Phi}$, $\beta$ and $\kappa$
are divergent at critical points, their divergence will be
cancelled in the Ehrenfest's equations. According to Eq. (15) and
(16), the condition as,

\begin{equation}
\frac{\triangle
C_{\phi}}{TQ}=\frac{\triangle\beta^{2}}{\triangle\kappa},
\end{equation}

\noindent is established in the background of static and
spherically symmetric spacetime with conformal anomaly. Here
$\triangle C_{\phi}=C_{\phi 2}-C_{\phi 1}$,
$\triangle\beta=\beta_{2}-\beta_{1}$ and
$\triangle\kappa=\kappa_{2}-\kappa_{1}$. Therefore, the black hole
with conformal anomaly can transform through the phase transition
point at $r_{c}$, and it is a second-order phase transition.

Both the expansivity and compressibility of Reissner-Nordstrom
black hole have discontinuities at the same position, which means
a phase transition may happen at $r_{+}=Q$. However, from phase 1
to phase 2 or conversely, the difference of the heat capacities
for the two kinds of phases respectively disappears $\triangle
C_{\Phi}=0$, but $\triangle\beta\neq 0$ and $\triangle\kappa\neq
0$. The necessary condition (23) is violated. The second order
phase transition of the black holes can not occur. It is
interesting that the Gauss-Bonnet term is fundamental for the
second order phase transition of the black holes. Only the black
holes with conformal anomaly can undergo to perform the second
order phase transition.

\vspace{0.8cm} \noindent \textbf{IV.\hspace{0.4cm}Conclusion}

In this work we discuss the phase transition in the black holes
with conformal anomaly in the Ehrenfest's scheme and further show
which kind of order the phase transition belongs to. The black
holes are thought as thermodynamic objects and their thermodynamic
quantities such as Hawking temperature, heat capacity at constant
potential, the analog of volume expansion coefficient and
isothermal compressibility are investigated from the first law of
black hole thermodynamics.

We find that the curves standing for the relationship between the
Hawking temperature and the mass of the black holes are
continuous, so the first order phase transition is excluded in
this description of evolution of black holes involving  conformal
anomaly. It is interesting that all of the heat capacity at
constant potential, the variables correspond volume expansivity
and isothermal compressibility are not continuous at phase
transition points and these variables satisfy the Ehrenfest's
equation. According to the plots and calculation, we find at the
points close to the critical point $r_{+}=r_{c}$, an unstable
black hole with higher mass and smaller radius changes into a
stable black hole with lower mass and larger radius. Moreover, we
prove analytically that the phase transition for the black holes
is the second order.

It is surprising that the heat capacity of black holes will
recover to be continuous and finite at any points if the influence
from conformal anomaly is omitted. One the other hand, the
Ehrenfest condition is broken for Reissner-Nordstrom black hole,
then the second-order phase transition will not emerge. In
conclusion, the conformal anomaly induces the second-order phase
transition of black holes.

\vspace{1cm}
\noindent \textbf{Acknowledge}

This work is supported by NSFC No. 10875043 and is partly
supported by the Shanghai Research Foundation No. 07dz22020.

\newpage
\begin{figure}
\setlength{\belowcaptionskip}{10pt} \centering
\includegraphics[width=15cm]{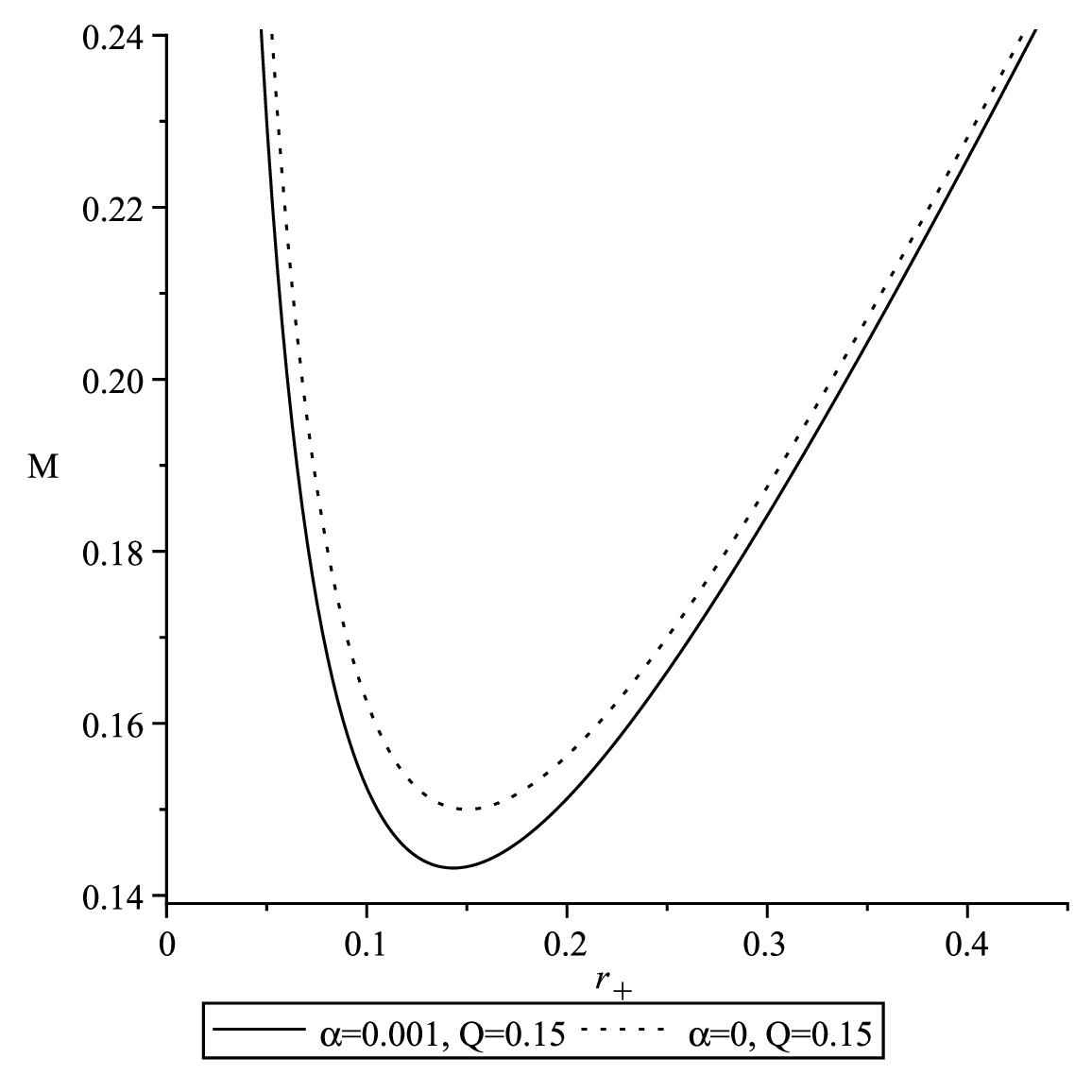}
\caption{The behaviour of black hole mass $M(r_{+})$ involving
conformal anomaly and a Reissner-Nordstrom black hole with respect
to the outer horizon of black hole. The solid and dotted lines
with $\alpha=0.001, 0$ respectively and $Q=0.15$.}
\end{figure}

\newpage
\begin{figure}
\setlength{\belowcaptionskip}{10pt} \centering
\includegraphics[width=15cm]{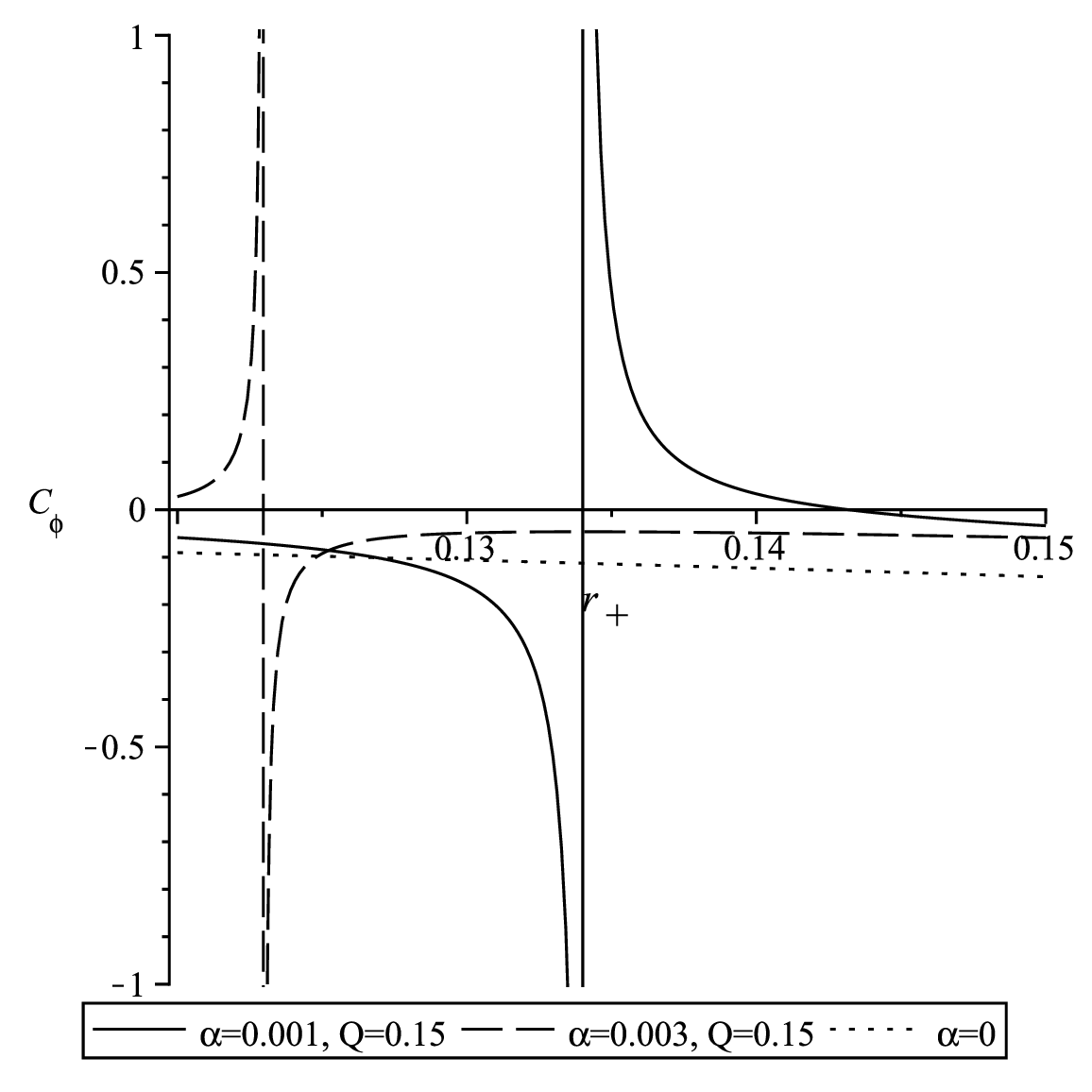}
\caption{The heat capacity of black holes as a function depends on
conformal anomaly. The solid line and the dotted line show the
behavior of the function with $\alpha =0.001$ and $\alpha =0.003$
respectively, while the dashed line corresponds the case of
Reissner-Nordstrom black hole with a vanished $\alpha$. All the
black holes contain a charge $Q=0.15$.}
\end{figure}

\newpage
\begin{figure}
\setlength{\belowcaptionskip}{10pt} \centering
  \includegraphics[width=15cm]{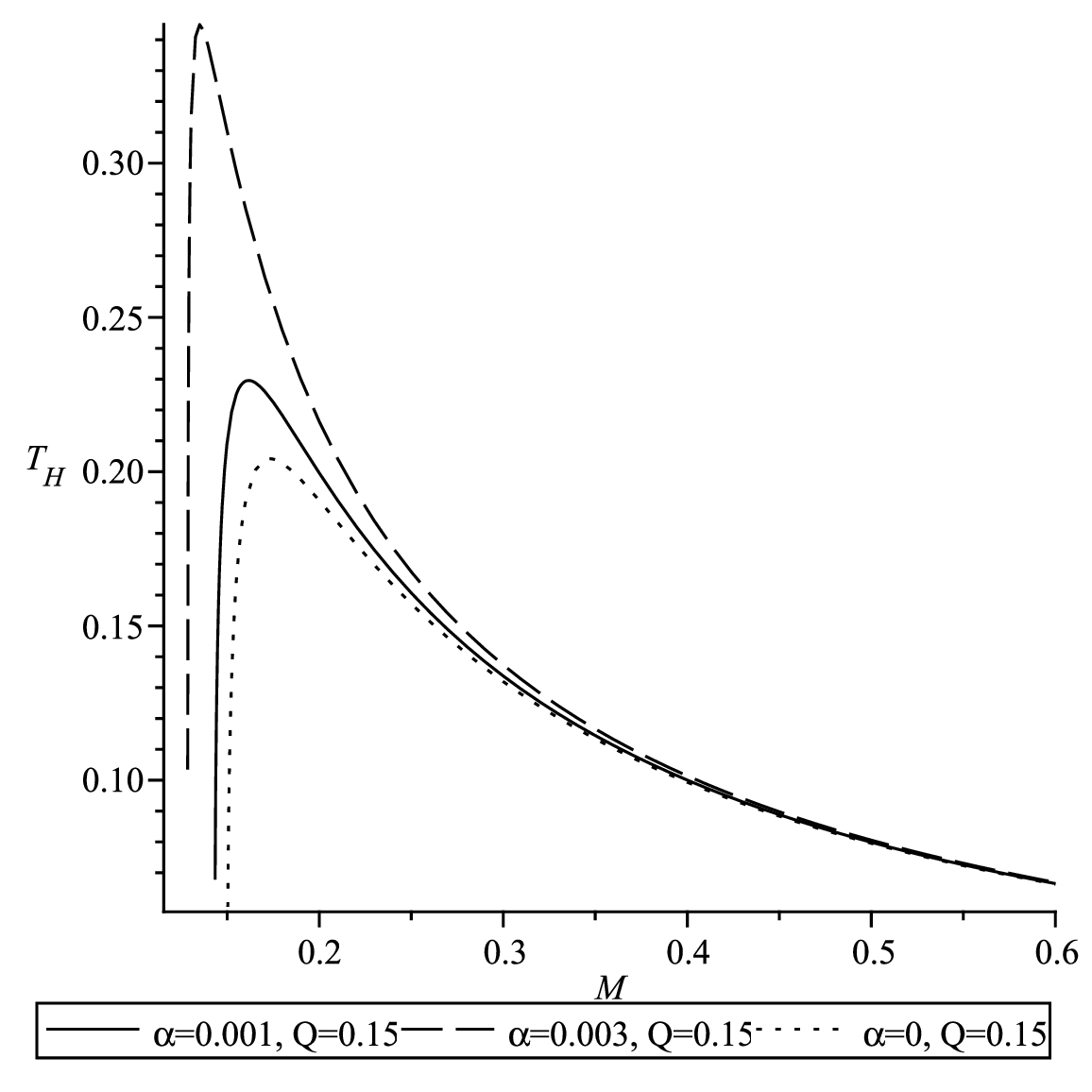}
  \caption{The relation between the Hawking temperature and the mass
  of a black hole with or without conformal anomaly.The solid,
  dashed and dotted lines denote ($\alpha, Q$) to be (0.001, 0.15), (0.003,
0.15) and (0,0.15) respectively.}
\end{figure}

\newpage
\begin{figure}
\setlength{\belowcaptionskip}{10pt} \centering
  \includegraphics[width=15cm]{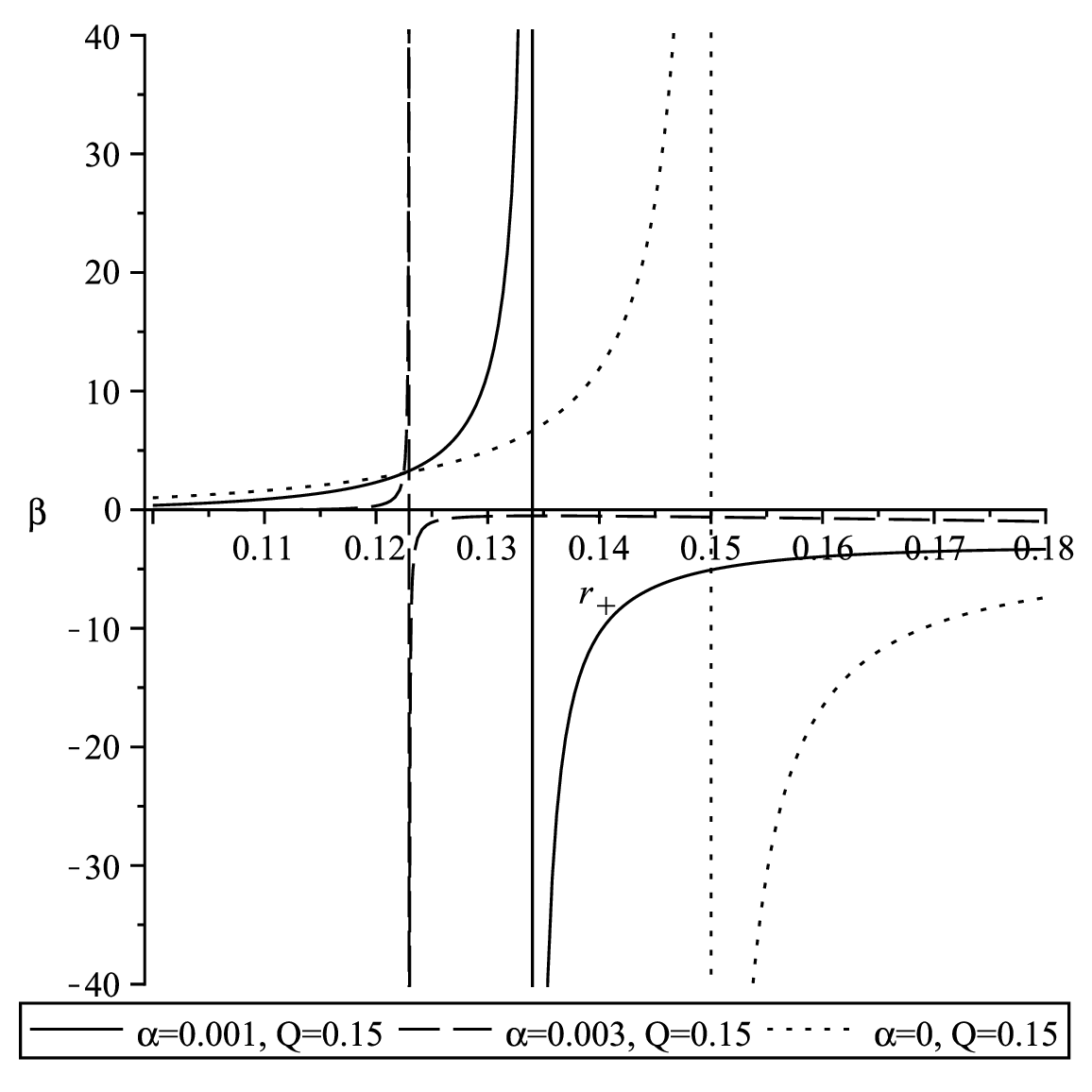}
  \caption{The dependence of the analog of volume expansion coefficient
  on the outer horizon for a black hole with different conformal anomaly $\alpha$.
  The solid, dashed and dotted lines denote $\alpha$ to be 0.001, 0.003, 0
  respectively and $Q=0.15$.}
\end{figure}

\newpage
\begin{figure}
\setlength{\belowcaptionskip}{10pt} \centering
  \includegraphics[width=15cm]{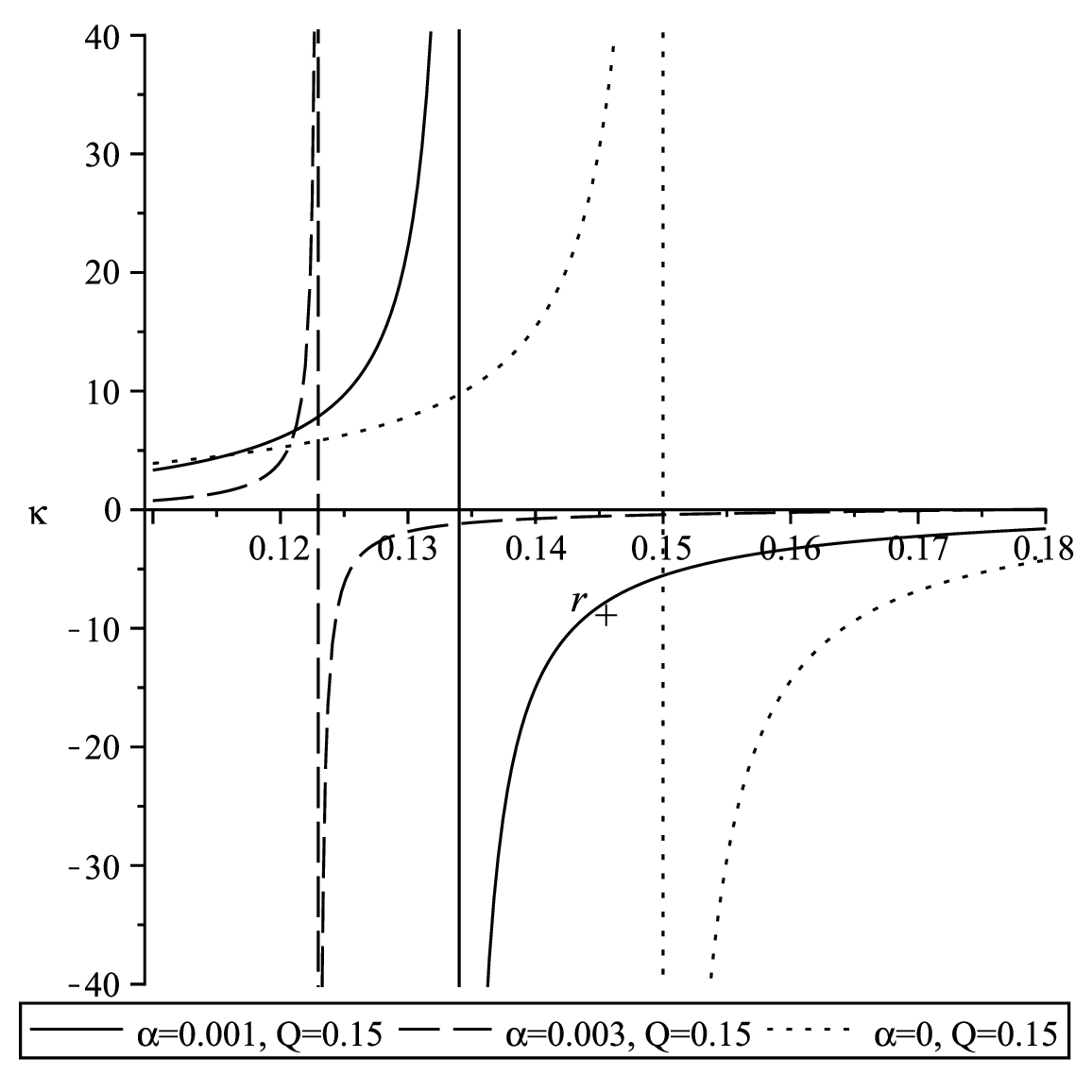}
  \caption{The dependence of the analog of isothermal compressibility
  on the outer horizon for a black hole with different conformal anomaly $\alpha$.
  The solid, dashed and dotted lines denote $\alpha$ to be 0.001, 0.003, 0
  respectively and $Q=0.15$.}
\end{figure}

\end{document}